\documentclass[10pt]{article}
\usepackage{graphicx}
\usepackage{booktabs}
\usepackage[english]{babel}
\usepackage{amsfonts,latexsym}
\usepackage{amsmath}
\usepackage{epstopdf} 
\usepackage{latexsym}
\usepackage[latin1]{inputenc}

\def\be{\begin{equation}}
\def\ee{\end{equation}}
\def\ba{\begin{eqnarray}}
\def\ea{\end{eqnarray}}
\def\parderiv#1#2{\frac{\partial #1}{\partial #2}}

\def\cwt{\cos(\omega t+\phi)}
\def\swt{\sin(\omega t+\phi)}

\begin{document}

\title{Comparison of Open and Solid Falling Retroreflector Gravimeters$^*$}

\author{Neil Ashby,$^{1,2}$  Derek van Westrum$^3$\\
$^1$Associate, Physical Measurement Laboratory,\\
 National Institute of Standards and Technology, Boulder, CO  80305\\
$^2$Department of Physics, University of Colorado, Boulder, CO 80309\\
 $^3$National Geodetic Survey, Boulder, CO 80305 USA}
\maketitle

\begin{abstract}
We study whether the optical properties of a solid glass retroreflector influence the value of the acceleration of gravity $g$ determined by dropping both solid and open retroreflectors in an absolute ballistic gravimeter.   The retroreflectors have equivalent optical centers and are dropped from the same height, at a fixed location, in the same gravimeter while recording time data corresponding to fixed fringe separation intervals of 400 fringes.  The data for both types of retroreflectors are processed with commercial software, as well as with independently developed software based on a relativistic treatment of the phase difference between reference beam and test beams, and a realistic treatment of the effect of frequency modulation, with modulation index $\beta \gg 1$, on the interference signal.  After applying corrections for polar motion, barometric admittance, tides, and ocean loading we find agreement between the values of $g$ determined with both types of retroreflectors, whether processed with commercial software or with our independently developed software.  We suggest two procedures for computing relativistic corrections; the two methods agree to better than .01 $\mu$Gal.
\footnote{This paper is a contribution of the U. S. government and is not subject to copyright. The use of commercial products does not imply endorsement by the U. S. Government.
}
\end{abstract}


\vspace{2pc}

\section{Introduction}

	In an absolute falling retroreflector gravimeter, a reference laser beam is partially reflected upwards from a stationary beamsplitter and then back down from a falling retroreflector (a ``cube"). Counting interference fringes between the reference and test beams may then be used to measure the local acceleration of gravity, $g$.  In a previous work\cite{ashby18}, the beams were assumed to combine at the beamsplitter; relativistic considerations led to the following expression for the interference signal at $Z=0$ (\cite{ashby18}, Eq. (33)):
\begin{multline}\label{finaltestphase}
\phi(t)=\frac{2( D n-d) \Omega}{c}
+\frac{2 \Omega Z_0}{c}
-\frac{2 (D n-d) \Omega V_0}{c^2}
-\frac{2 \Omega Z_0 V_0}{c^2}\\
{}+t(-\Omega +\frac{2\Omega V_0}{c}-\frac{2\Omega V_0^2}{c^2}+\frac{2\Omega g ( D n-d)}{c^2}+\frac{2 g \Omega Z_0}{c^2})\\
+t^2(-\frac{g \Omega}{c}+\frac{3 g \Omega V_0}{c^2})-\frac{g^2 \Omega t^3}{c^2}\\
{}+\gamma \Omega \bigg( t(-\frac{2 ( D n-d)Z_0}{c^2}-\frac{2 Z_0^2}{c^2})
+t^2(\frac{Z_0}{c}-\frac{(D n-d) V_0}{c^2}-\frac{4V_0Z_0}{c^2})\\
{}+t^3(\frac{V_0}{3c}
+\frac{g( D n-d)}{ 3c^2}-\frac{4 V_0^2}{3 c^2}
+\frac{7g Z_0}{3c^2})
{}+t^4(-\frac{g}{12 c}+\frac{5 g V_0}{4 c^2})-\frac{g^2 t^5}{4 c^2}  \bigg)
\,,
\end{multline}	
where $c$ is the speed of light, $\Omega$ is angular frequency of the light, $D$ is the face-to-corner distance of the retroreflector (the "cube"), $d$ is the distance from face to center of mass, $n$ is the index of refraction, $Z_0$ and $V_0$ are the position and velocity of the cube at the initial time $T=0$, $g$ is the acceleration of gravity at the beamsplitter and $\gamma$ is the gravity gradient.  The origin of coordinates is at the beamsplitter and the coordinate $Z$ is positive upwards.

Eq. (\ref{finaltestphase}) is the basis for the present study. The apparently large effect on $g$ arising from dimensions and refractive index of the cube resulted in several critical comments\cite{kren18,nagornyi18,svitlov18}, and replies\cite{ashby18k,ashby18n,ashby18s}, without throwing light on the real issues. 
	
	There are two problems with the literal application of an interference signal that is produced at the beamsplitter.  The first is that in some gravimeters the reference beam, or the test beam, or both, are transported unequal distances by fibers and/or mirrors before interference takes place\cite{niebauer95}; this introduces an unknown constant phase between the reference and test beams.  Also, the interference signal occurs within the argument of a trigonometric function--such as a cosine--and is undetermined to within a constant added integral multiple of $2\pi$.  This issue is treated in detail in Sect. 3, where a theory of frequency modulation by a signal with modulation index $\beta$, which is large compared to unity, is developed.  These facts lead to reconsideration of the choice of fit parameters and the effect on $g$ of the cube's properties. 
	
In the following section, we discuss the choice of fit parameters, and the effect on $g$.  The constant parameter combination $Z_0+Dn-d$ disappears from the non-relativistic part of the phase and becomes a small relativistic contribution.  Two methods are presented in Sect. 4 for explicitly computing relativistic effects arising from the cube's velocity.  Measurements made at Table Mountain, Boulder, CO, with open and solid cubes of comparable dimensions, in the same dropping chamber, are described and discussed in Sect. 5; to within measurement uncertainties the same value of $g$ was obtained--at the same location and during comparable time periods.  The relativistic corrections--obtained in two ways--were in agreement to within $0.01\ \mu$Gal.\footnote{$1\ \mu {\rm Gal}=10^{-10}{\rm\ m\ s}^{-2}$}  Sects. 6 and 7 compare the data analysis using commercial software and software based on the present theory, respectively.

\section{The parameter $Z_0$}
	 Four constant parameters are usually used to describe the falling cube:  The initial position $Z_0$, initial velocity $V_0$, and the acceleration of gravity $g$ and its gradient $\gamma$ at the chosen origin of coordinates, which in \cite{ashby18} was the point of separation and recombination of the beams, at the beamsplitter.  The interference phase at the beamsplitter was calculated neglecting quadratic and higher order terms in the gravity gradient $\gamma$ and to this order may be compactly expressed as
\be\label{phasebs}
\phi(t)=\Omega F(t)=-\frac{2\Omega}{c}(Z_{cm}(t)+D n-d)\bigg(1-\frac{V_{cm}(t)}{c}\bigg)\,,
\ee
$F(t)$ is independent of frequency, and can be separated into non-relativistic and relativistic parts:
\be\label{separation}
F(t)= \frac{1}{c} F_{nr}(t)+\frac{1}{c^2}F_{rel}(t)\,.
\ee 
$Z_{cm}(t)$ and $V_{cm}(t)$ are the center of mass (CM) position and velocity, given by
\be\label{Zcm}
Z_{cm}(t)=Z_0+V_0 t-\frac{g t^2}{2}+\gamma\bigg(\frac{Z_0 t^2}{2}+\frac{V_0 t^3}{6}-\frac{g t^4}{24}\bigg) \,;
\ee
\be\label{Vcm}
V_{cm}(t)=V_0-g t+\gamma\bigg(Z_0 t+\frac{V_0 t^2}{2}-\frac{g t^3}{6}\bigg)\,.
\ee
Eq. (\ref{phasebs}) is mathematically equivalent, to first order in $\gamma$, to the interference phase, Eq. (\ref{finaltestphase}), and allows one to cleanly separate relativistic from non-relativistic contributions.  The purpose of this section is to show that, contrary to many statements in the literature, $Z_0$ cannot always  be considered a useful parameter in datafitting. To make this clearer, in the remainder of this section we discuss a non-relativistic model,  without the complication of relativistic effects.

  Assuming there is no modulation, the phase of the interference signal at the beamsplitter where recombination was assumed to occur is
\ba
\phi(t)= -\frac{2 \Omega}{c}\bigg( Z_0+Dn-d+V_0 t-\frac{1}{2}g t^2+\gamma(\frac{1}{2}Z_0 t^2+\frac{1}{6}V_0 t^3-\frac{1}{24} g t^4 \bigg)\notag\\
\times \bigg(1-\frac{V_{cm}(t)}{c}\bigg)\,.\hbox to 2 truein{} 
\ea
The nonrelativistic form of this limiting phase is obtained by making the replacement 
\be
\Omega \rightarrow\frac{ 2 \pi c}{\lambda}\,,
\ee
and neglecting $V_{cm}/c$.  Then in the non-relativistic approximation,
\be\label{phase0}
\phi(t)
=-\frac{4\pi}{\lambda}\bigg( Z_0+Dn-d+V_0 t-\frac{1}{2}g t^2+\gamma\big(\frac{1}{2}Z_0 t^2+\frac{1}{6}V_0 t^3-\frac{1}{24} g t^4 \big) \bigg)\,. 
\ee
A minus sign was introduced so that the coefficient of the term in $g t^2$ is positive, corresponding to measurements in which the number of fringes increases with time.  This phase occurs in the argument of a cosine function, to which an arbitrary multiple of $2\pi$ can be added without changing the cosine's magnitude; this is discussed in detail in Sect. 3.  The observable phase is equivalent to
\be
\phi(t)+2\pi M\,,
\ee
where $M$ is an unknown, arbitrary integer.  Any constant contribution on the right side of Eq. (\ref{phase0}) can be combined with the term $2\pi M$ leaving a residual phase $\psi$:
\be 
2\pi M-\frac{4\pi}{\lambda}(Z_0+D n -d)=\psi\,.
\ee
The observable phase in a non-relativistic approximation is therefore
\be\label{phase1} 
\phi(t)=\psi-\frac{4\pi}{\lambda}\bigg(V_0 t-\frac{1}{2}g t^2+\gamma(\frac{1}{2}Z_0 t^2+\frac{1}{6}V_0 t^3-\frac{1}{24} g t^4) \bigg)\,.
\ee
The terms $Dn-d$ contribute instead to the relativistic part of the phase. The parameter $Z_0$ now contributes only weakly, since it is multiplied by the gradient $\gamma$. The phase in Eq. (\ref{phase1}) suggests fitting the data (neglecting relativistic effects) using the parameters $\{\psi,Z_0,V_0, g\}$.  The results are not reliable because both $g$ and $Z_0$ have coefficients proportional to $t^2$: $g$ and $Z_0$ are highly correlated and the covariance matrix is nearly singular.  	Inclusion of relativistic terms does not improve the situation. 

The difficulty may be remedied by combining $g$ and $Z_0$ into a single parameter.  Let
\be\label{defofgtop}
g_t=g-\gamma Z_0\,.
\ee
We also replace $g$ in the last term of Eq. (\ref{phase1}) by $g_t$.  At a maximum drop time of 0.3 s., with $Z_0 \approx .75 {\rm\ m}$ the error of the latter replacement is of order
\be 
\frac{\pi}{6\lambda}\gamma^2 Z_0 t^4\approx 5 \times 10^{-7}   {\rm\ radians},
\ee
which is negligible as is cannot give rise to a significant number of fringes. Then $Z_0$ disappears from the non-relativistic model.  The fit parameters become $\psi,V_0,g_t$. In this scheme the covariance matrix is non-singular.  Relativistic contributions depend on the velocity of the center of mass, which can also be written to the same order in terms of $g_t$:
\ba
V_{cm}(t)=V_0-gt+\gamma(Z_0 t+\frac{1}{2} V_0 t^2-\frac{1}{6} g t^3)\notag\\
\approx V_0-g_t t+\gamma(\frac{1}{2} V_0 t^2-\frac{1}{6} g_t t^3)\,. 
\ea
 The phase error introduced by the quadratic term in $\gamma$ is less than $10^{-20}$ radians.  However the value of $g_t$ at the ``top of the drop" is no longer at a fixed location in the chosen coordinate system, but includes variations in initial drop position, thus might suffer from greater errors than if it were  evaluated at a fixed location in the coordinate system.  This is consistent with some previously published models (see e.g., Appendix of \cite{niebauer95}), in which the acceleration of the cube is written as 
\be 
\ddot Z =-g+\gamma(Z-Z_0)\,.
\ee
Here $g$ would be evaluated for a particular drop at the initial instant when $Z=Z_0$ and might vary depending on the consistency of release. 
	
\section{Modulation; choice of fit parameters}

The purpose of this section is to develop an appropriate expression for the frequency-modulated interference signal with large modulation index, including relativistic effects.   This is based on the interference phase in Eq. (\ref{phasebs}).  If the reference signal is modulated at angular frequency $\omega$ with amplitude $ \omega_m $, the instantaneous angular frequency can be written $\Omega +\omega_m \cos(\omega t +\phi)$.  Then apart from a constant of integration, the signal phase will be 
\be  
\int^t\big(\Omega +\omega_m \cos(\omega t+\phi) \big) dt=\Omega t+\beta \sin(\omega t+\phi)\,,
\ee
where $ \beta=\omega_m/\omega $ is the modulation index.  In the gravimeter used in the present experiment, the laser light is modulated at $\omega=(2 \pi) \times 8333$ Hz with peak-to-peak amplitude 6 MHz, so
\be 
\beta \approx \frac{3 \times 10^6}{8333}=340 \gg 1 \,.
\ee
An expansion for small modulation index is inappropriate.  Let the reference beam emerging from the beam splitter be represented by the scalar wavefunction
\be
\psi_{ref}=e^{-i \Omega t-i \beta \swt}\,,
\ee
where $\beta \sin(\phi)$ represents the phase difference between the reference beam and the modulation at time $t=0$.     

Thus, in addition to the principal frequency $\Omega$, frequencies $\Omega \pm n\omega$ will be present, where $n$ can be any integer.  For a signal of frequency $\Omega'$, Eq. (\ref{phasebs}) shows that the final phase of the signal that passes up through the retroreflector and back to the beamsplitter is proportional to $\Omega'$, so that at the beamsplitter the component of frequency $\Omega$ becomes
\be 
e^{-i\Omega t}\rightarrow e^{-i\Omega(t+F(t))};
\ee
and for the side frequency terms,
\be 
e^{-i\Omega t \pm i n \omega t}\rightarrow e^{(-i \Omega \pm i n \omega)(t+F(t))}\,.
\ee
The function $F(t)$ was defined in Eq. (\ref{separation}).  The test beam amplitude is therefore
\be
\psi_{test} = e^{-i\Omega(t+F(t)+\beta \sin(\omega t+\omega F(t)+\phi)}\,.
\ee
We assume the wavefunctions are superimposed at the beamsplitter with equal amplitudes, so the net signal is
\be 
\psi=\psi_{ref}+\psi_{test}\,.
\ee
The intensity of the signal will be proportional to 
\begin{multline}
\vert \psi \vert^2=\vert \psi_{ref} \vert^2+\vert \psi_{test} \vert^2+\psi_{ref}\psi_{test}^*+\psi_{ref}^*\psi_{test}  \\
=2+2 \cos(\Omega F(t)+\beta \big( \sin(\omega t+\phi+ \omega F(t))-\sin(\omega t+\phi)\big)\\
=2+2 \cos(\Omega F(t)+\beta \big(\sin(\omega t+\phi )(\cos(\omega F(t))-1)+\cos(\omega t+\phi)\sin(\omega F(t)\big)\,.
\end{multline}
In the present experiment, for each drop 2700 times were recorded separated by 400 fringes, so the maximum value of the function $F(t)$ can be estimated by setting
\be 
 F(t_{max})\Omega = (2 \pi) \times 400 \times 2700, \quad{\rm or} \quad F_{max} = 2.85 \times 10^{-9} {\rm s}\,;
\ee
then $\omega F_{max} \approx 0.00015 \ll 1 $.  Even with $\beta$ as large as it is, we can make the approximations
\be 
\cos(\omega F(t))\approx 1,\quad \sin(\omega F(t)) \approx \omega F(t)\,.
\ee
The signal intensity then simplifies to
\be 
\vert \psi \vert^2=2+2 \cos(\Omega F(t) + \beta \omega F(t) \cos(\omega t+\phi) )\,.
\ee
The function $F(t)$ is given in Eq. (\ref{separation}) as the sum of non-relativistic and relativistic contributions:  
\be\label{eq26}
F_{nr}(t)= -\frac{2}{c}(Z_{cm}(t)+D n -d);\quad F_{rel}=\frac{2 V_{cm}(t)}{c^2}(Z_{cm}(t)+Dn-d)\,.
\ee 
one factor $1/c$ is cancelled out when the replacement $\Omega\rightarrow (2\pi c/\lambda)$ is made.  The number of fringes counted $N(t)$ will be determined by the argument of the interference cosine, which is
\be
N_{total}=\frac{1}{2\pi}\bigg(\Omega F(t)+\beta\big(\cos(\omega t+\phi) \omega F(t) \big)\bigg)\,.
\ee
The phase is undetermined to within an added integral multiple of $2\pi$:
\begin{multline}\label{fringeswithmod}
N(t)=\frac{1}{2\pi}\big(2 \pi M+\Omega F(t)+\beta\cwt \omega F(t)\big)\\
=M+\frac{1}{2\pi}\bigg(\frac{2\pi c}{\lambda}\bigg(\frac{1}{c}F_{nr}+\frac{1}{c^2}F_{rel}\bigg)\hbox to 1.5 truein{}\\
\hbox to .6 truein{}+\beta \omega (\cos(\phi)\cos(\omega t)-\sin(\phi)\sin(\omega t))\bigg(\frac{1}{c}F_{nr}+\frac{1}
 {c^2}F_{rel}\bigg)\bigg)\\
 =M+\frac{1}{\lambda}\big(1+\lambda A \cos(\omega t)+\lambda B \sin(\omega t)\big)\bigg(F_{nr}+\frac{1}{c} F_{rel}\bigg)\,,\hbox to .35 truein{}
\end{multline}
where $M$ is an unknown integer and
\be 
A\cos(\omega t)+B\sin(\omega t)=\frac{\beta \omega}{2\pi c}\cwt\,.
\ee
Thus the modulation index is related to the amplitudes $A$ and $B$ by
\be 
\sqrt{A^2+B^2}=\frac{ \beta \omega }{2 \pi c}\,;
\ee
although the modulation index is large, the coefficients $A$ and $B$ will be small.
Apart from the arbitrary whole number $M$, The nonrelativistic fringe function is then
\be\label{nrfringenumber}
N_{nr}(t)=-2\bigg(\frac{1}{\lambda}+\big(A\cos(\omega t)+B\sin(\omega t)\big)\bigg)(Z_{cm}(t)+d n-d)
\ee
and the relativistic contributions, showing explicitly the factor $c^{-1}$, are
\be\label{relfringefunction}
N_{rel}(t)=N(t)-N_{nr}(t)=\frac{1}{c \lambda}F_{rel}(t)+\frac{1}{c}\big(A\cos(\omega t)+B\sin(\omega t)\big) F_{rel}(t)\,.
\ee
The functions $F_{nr}$ and $F_{rel}$ are given by
\be\label{fringefunctions}
F_{nr}+\frac{1}{c} F_{rel}=-2\big(Z_{cm}(t)+D n-d\big)\bigg(1-\frac{V_{cm}(t)}{c}\bigg)\,.
\ee
If $\omega F(t)$, which is slowly varying, were treated as a constant then we would simply have the two parameters $A$ and $B$ to add to the list of parameters to be fit.  Including two such constant parameters typically reduces the residuals by a factor of 5 or more without having much impact on $g$.  Including the variation due to $F(t)$ results in reduction in the residuals by roughly 3\% to 4\% for this experiment.

 As discussed in the previous section, writing the CM position and velocity in terms of $g_t$, eliminates $Z_0$ from the model. Consider the non-relativistic contributions to the fringe number in Eq. (\ref{nrfringenumber}) at $t=0$, where we regard the arbitrary $M$ as part of the non-relativistic contribution.  (Our approach is to group the relativistic constant contributions with the relativistic corrections.) 
\be 
N_{nr}(0)=M+\frac{1}{\lambda}(1+\lambda A)(Z_0+Dn-d)\,.
\ee
We replace these constant contributions by a residual phase constant $\frac{\psi}{2\pi}$, so the number of observed fringes becomes
\ba\label{finalmodel} 
N(t)=\frac{\psi}{2\pi}-\frac{1}{\lambda}(1+\lambda A)(Z_0+Dn-d)\notag\hbox to 1 truein{}\\
+\frac{1}{\lambda}\big(1+\lambda A \cos(\omega t)+\lambda B \sin(\omega t)\big)\bigg(F_{nr}+\frac{1}{c} F_{rel}\bigg)\,.
\ea
As discussed in Sect. 2, introducing $g_t$ as in Eq. (\ref{defofgtop})  and making the approximation $ g\rightarrow g_t$ removes the parameter $Z_0$ from the model.  The parameters to be fit are $\psi, V_0, g_t, A, B$.  We discuss fitting data obtained with both open and solid retroreflectors using this model in Sects. 5 and 6.

\section{Relativistic corrections}

  Fitting a theoretical expression for fringe number $N(t_j)$ that depends on a set of fit parameters $p_k$ involves the covariance matrix
\be
C_{kl}=\sum_j \parderiv{N(t_j)}{p_k} \parderiv{N(t_j)}{p_l}\,,
\ee
where the recorded times are $t_j$ at epoch $j$ when the measured total number of fringes is $M_j=M(t_j)$.  Guessing values $p_k^{(0)}$ for the fit parameters, and substituting these into the $N(t_j)$ and its partial derivatives, the next approximation will be
\be\label{iterate} 
p_k^{(1)}=p_k^{(0)}+\sum_l \big(C^{-1}\big)_{kl}\sum_j(M_j-N(t_j)) \parderiv{N(t_j)}{p_l}\,.
\ee
The improved parameter values $p_k^{(1)}$ are substituted into the right side of Eq. (\ref{iterate}) and the process is repeated until there is no further change in any parameter.  At this point the covariance matrix is non-singular so the residuals satisfy
\be\label{residuals}
\sum_j(M_j-N(t_j)) \parderiv{N(t_j)}{p_l}=0\,.
\ee
Eq. (\ref{residuals}) permits the development of an explicit expression for the relativistic contribution to the determination of $g$.  Divide the theoretical expression for the fringe number into non-relativistic and relativistic contributions; the latter will be of order $c^{-1}$, where $c$ is the speed of light.  Only first order terms in an expansion in powers of $c^{-1}$ need be kept.  Then
\be\label{Nexpansion}
N(t_j)=N_{nr}(t_j)+N_{rel}(t_j),
\ee
and the covariance matrix becomes
\ba
C_{kl}=\sum_j \parderiv{N_{nr}}{p_k} \parderiv{N_{nr}}{p_l}+\sum_j\bigg( \parderiv{N_{nr}}{p_k} \parderiv{N_{rel}}{p_l}+ \parderiv{N_{rel}}{p_k} \parderiv{N_{nr}}{p_l}\bigg)\\
=  C_{kl}^{nr}+C_{kl}^{rel}\,,\hbox  to 2 truein {}
\ea
where to simplify the notation the arguments $t_j$ have been suppressed.  The inverse covariance matrix can be expanded in powers of $C^{-1}$:
\be 
C^{-1}=(C^{nr})^{-1}-(C^{nr})^{-1}C^{rel}(C^{nr})^{-1}\,.
\ee
Then Eq. (\ref{iterate}) becomes approximately
\ba\label{iteratea} 
p_k^{(1)}=p_k^{(0)}+\sum_l (C^{nr})_{kl}^{-1}
\sum_j (M_j-N_{nr}(t_j)) \parderiv{N_{nr}(t_j)}{p_l}\nonumber\\
+\sum_l \bigg((C^{nr})^{-1}C^{rel}(C^{nr})^{-1}\bigg)_{kl} \sum_j (M_j-N_{nr}(t_j)) \parderiv{N_{nr}(t_j)}{p_l}\nonumber\\
+\sum_l(C^{nr})^{-1}_{kl}\sum_j(M_j-N_{nr}(t_j)) \parderiv{N_{rel}(t_j)}{p_l}\nonumber\\
-\sum_l(C^{nr})^{-1}_{kl} \sum_j N_{rel}(t_j)\parderiv{N_{nr}(t_j)}{p_l}\,.
\ea
Then one can consider the following approximate iteration scheme.  First neglect the last three lines of Eq. (\ref{iteratea}), fitting the parameters using only the non-relativistic theory expressed in the first line.  The non-relativistic values of these parameters can then be inserted into the remaining terms in lines 2 through 4 since each of these terms already contains a factor $c^{-1}$.  But the iteration process has converged and reduces the first line,
\be\label{relcorr}
\sum_l (C^{nr})^{-1}_{kl}\sum_j (M_j-N_{nr}(t_j)) \parderiv{N_{nr}(t_j)}{p_l}\nonumber\,,
\ee
to a negligible level.  Then line 2 is also negligible.  Line 3 is not identically zero but is found to be negligibly small because each term $M_j-N_{nr}(t_j)$ is small, a result of the fit.  The relativistic contributions to the parameters are then simply:
\be\label{relativisticcorrections}
p_k^{rel}=-\sum_l(C^{nr})^{-1}_{kl} \sum_j N_{rel}(t_j)\parderiv{N_{nr}(t_j)}{p_l}\,.
\ee
One may check Eq. (\ref{relcorr}) by first fitting the parameters using Eq. (\ref{iterate}) with the complete expression for $N(t_j)$, instead of $N_{nr}$, and then subtracting the non-relativistic fit obtained from iterating the non-relativistic model:
\be\label{nonreliterate} 
(p_k^{nr})^{(1)}=(p_k^{nr})^{(0)}+\sum_l \big((C^{nr})^{-1}\big)_{kl}\sum_j(M_j-N_{nr}(t_j)) \parderiv{N{nr}(t_j)}{p_l}\,.
\ee
The relativistic corrections, Eq. ({\ref{relcorr}), were found to agree within .05 $\mu$Gal with the differences in the fits with open and solid retroreflectors reported in Tables 1 through 4.

\subsection{ Correcting recorded times for relativity.}  Another approach is to model the fringe count non-relativistically, as in Eq. (\ref{nrfringenumber}), but seek corrections $\delta t$ to the recorded times, such that correcting the time data brings the data into agreement with the full relativistic model.  The nonrelativistic fringe function can be written:
\be\label{nonrelmodel}
N_{nr}(t)=-\frac{2}{\lambda}\big(Z_{cm}(t)+ D n-d\big)\bigg(1+\frac{\beta \omega \lambda}{2 \pi c}\cos(\omega t+\phi)\bigg)\,,
\ee
and the relativistic contribution to the fringe function, from Eqs. (\ref{relfringefunction}) and (\ref{fringefunctions}) is
\be
N_{rel}=\frac{2 V_{cm}(t)}{\lambda c}\big(Z_{cm}(t)+ D n-d\big)\big(1+\frac{\beta\omega\lambda}{2\pi c}\cos(\omega t+\phi) \big)\,.
\ee
We replace $t$ by $t+\delta t$ in Eq. (\ref{nonrelmodel}) and expand to first order:
\ba\label{expansion}
N_{nr}(t+\delta t) = N_{nr}(t)+\frac{d N_{nr}(t)}{dt} \delta t\nonumber\\
= N_{nr}(t)-\frac{2}{\lambda} V_{cm}(t) \bigg(1+\frac{\beta \omega \lambda}{2 \pi c}\cos(\omega t+\phi)\bigg)\delta t\nonumber\\
+2\big(Z_{cm}(t)+ D n -d\big)\big(\frac{\beta \omega^2}{2 \pi c}\sin(\omega t+\phi) \big)\delta t\,.
\ea
The extra terms in $\delta t$ will reproduce the relativistic contributions provided that
\be\label{deltat}
\delta t=-\frac{1}{c}(Z_{cm}(t)+ D  n-d)\,,
\ee
and provided the extra term in $\omega^2$, which arose from the time derivative of the modulation, is negligible.  But substituting the solution Eq. (\ref{deltat}) into the last line on Eq. (\ref{expansion})
gives a coefficient that is extremely small:
\be 
\frac{\beta \omega^2}{\pi c^2} \approx 3 \times 10^{-6}\,.
\ee
This is a negligible contribution to the fringe count. 

If it were not for modulation terms, the time correction could be expressed as
\be 
\delta t =\frac{\lambda}{2 c}\big(N_{nr}(t)-\frac{\psi}{2\pi}\big)+\frac{1}{2 c}(Z_0+Dn-d)\,,
\ee
where $N_{nr}(t)$ is the observed number of fringes at the recorded time $t$.  A similar correction is used in the g9 software.

\section{Experimental Setup}

A Micro-g Lacoste FG5-X absolute gravimeter\cite{niebauer95,niebauer12}, S/N 302,\footnote{The U.S. Government does not endorse any particular equipment.}  was operated at the NOAA Table Mountain Geophysical Observatory in August of 2019.\cite{vanWestrum19}.  Over the course of three days, two different dropping chambers were installed on the instrument.  Chamber \#1 is a ``standard" FG5-X dropping chamber, (S/N 102), which employs a solid glass, ``closed face" (n = 1.5) retroreflector.   Chamber \#2 is a research unit, similar in construction to \#1, but with a hollow ``open face" retroreflector comprised of three mirrors.  In both chambers, the retroreflectors are housed in a ``test mass" which undergoes free-fall acceleration.  
Because the center of mass of each test mass needs to be coincident with the optical center of the retroreflector (to minimize second order rotation effects), the test mass of Chamber \#2 is necessarily of slightly different construction than that of \#1 (Figure 1).  Otherwise, the dropping chambers are 
\begin{figure}[t]\label{fig1}
\centering
\includegraphics[width=4.75 truein]{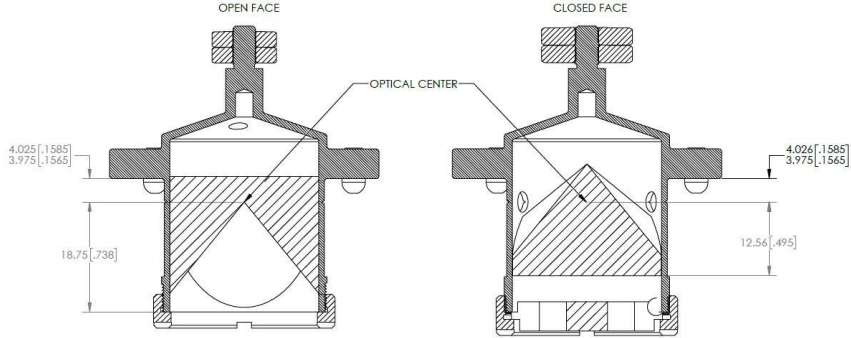}
\caption{The two test masses used in the experiment.  The object on the left has an ``open face" retroreflector (Chamber \#2), while the object on the right has the standard ``closed face" retroreflector (Chamber \#1). Diagram courtesy of Micro-g Lacoste.}
\end{figure}
identical:  same counter mass design, drop length, nominal vacuum level, etc.  
The rest of FG5X 302's hardware - the laser, interferometer, the seismic isolation device
 (``Superspring"), the data acquisition hardware, and all cabling - was used in all set ups to minimize any systematic effects between the observations.  Before the observations, both chambers were tested to quantify any horizontal velocity of the test mass at the beginning of the drop.   Any such motion in the east-west direction will result in a Coriolis acceleration that could systematically and significantly  bias the gravity results.  However, both chambers registered negligible velocities:  Chamber \#1 has a  Coriolis acceleration equivalent to $0.2 \pm 0.1\ \mu$Gal, and Chamber \#2 has no noticeable acceleration at all: $ 0.0 \pm 0.1\ \mu$Gal.  Both values are well within the total uncertainty of the final gravity values, and the ``Coriolis Effect" is henceforth neglected in the analysis.
 
 The first observations employed Chamber \#1 and comprised 12 sets of 100 drops at a 5 second drop interval, for a total acquisition time of 12 hours.  A second data set with Chamber \#1 was then acquired two days later, with a single set of 1000 drops (84 minutes).  Next, Chamber \#2 was installed, and a set of 500 drops was then acquired (42 minutes).   This process was repeated two more times (500 drop sets) for a total of three iterations between the chambers.
 
 Between each setup, care was taken to realign the system to the local plumb line, maximize the interferometer fringe contrast (always between 390 mV and 400 mV, peak-to-peak), and accurately measure the height of the dropping chamber.  The g9 software of Micro-g LaCoste was used to process the data, and the default corrections for Earth tide, ocean loading, barometric admittance, and polar motion were applied on a drop by drop basis.
 
Regardless of which chamber was installed, the drop to drop standard deviation for all runs was between 1.9 and 2.1 $\mu$Gal.  These small values are a strong indication that the object was not significantly rotating in either chamber.  Such an effect is usually sporadic, and if it had been present, it would have led to an increased scatter in the results.

\section{Experimental Results I: Applying Commercial Software}

The observation details and gravity results are listed in Table 1.   The slight difference in the starting heights of the test masses is accounted for by transferring the $g$ value from the actual ``top of the drop," to a common height of 139.6 cm above the bench mark (an approximate average of the various raw heights).  A vertical gravity gradient of -3.18 $\mu$Gal/cm observed at the TMGO Pier, AT, was used to make this transfer.  The listed ``uncertainty" is simply the total uncertainty that is reported by the g9 software.   As discussed below, it will be reduced when we calculate the difference in gravity observed by the two chambers.
\begin{table}[h]
\caption{Experimental Results using commercial software.  Height is the distance from the bench mark to the optical center of the retroreflector at the top of the drop.   The gravity values are differenced from 979 622 000 $\mu$Gal.  The vertical gravity gradient is used to transfer the raw values to a common height of 139.6 cm above the bench mark. Chamber \# 1 is the solid retroreflector.}
\vskip 8 pt
\begin{center}
\begin{tabular}{c c c c c c}
\toprule
Date & Chamber & Number & Height& Gravity & Uncertainty \\
& & of Drops & (cm)&  ($\mu$Gal) &\\
\midrule
8.6.2019 & \# 1 & 1200 & 139.8 & 710.7 & 1.8\\
8.8.2019 & \# 1 & 1000 & 139.8 & 709.9 & 1.8\\
8.8.2019 & \# 2 & 500 & 139.45 & 711.3 & 1.8\\
8.9.2019 & \# 1 & 500 & 139.75 & 710.8 & 1.8\\
8.9.2019 & \# 2 & 500 & 139.5  & 709.9 & 1.8\\
8.9.2019 & \# 1 & 500 & 139.8  & 710.0 & 1.8\\
8.9.2019 & \# 2 & 500 & 139.45  & 711.4 & 1.8\\
\bottomrule
\end{tabular}
\end{center}
\end{table}

Table 2 lists the mean value of gravity from each dropping chamber, along with the standard deviation of the observations and the standard statistical error.  The systematic uncertainty for each measurement is estimated to be 0.5 $\mu$Gal.  This was determined by using the default systematic uncertainty values in the g9 software, but setting the uncertainties of the laser, clock, ``system," and barometer to zero (as these are common mode for all of the observations here).  The ``setup" uncertainty was estimated to be 0.5 $\mu$Gal (primarily attributed to imperfections in the plumb line alignment and height determination).  Next, the statistical error, $\sigma/\sqrt{N}$, is added in quadrature with the systematic to get a total uncertainty of about 1 $\mu$Gal for each chamber (compare with the default uncertainty of 1.8 $\mu$Gal reported in Table 1).  The final difference in gravity as measured by the two chambers (glass - open) is  -0.5 $\pm$ 1.25 $\mu$Gal, where the final uncertainty on the (correlated) difference is the sum of the combined uncertainties.

\begin{table}[h]
\caption{Experimental results using commercial software. The mean gravity values are differenced from 979\ 622\ 000 $\mu$Gal.  All units are $\mu$Gal.}
\vskip 8 pt
\begin{center}
\begin{tabular}{c c}
\toprule
Mean Gravity Chamber \# 1 (Glass) & 710.4 \\
\midrule
Error of the Mean & 1.0\\
Systematic Uncertainty &0.5 \\
 Combined Uncertainty & 1.1\\

\midrule
Mean Gravity Chamber \# 2 (Open) & 710.9 \\
\midrule
Error of the Mean &1.2\\
Systematic Uncertainty &0.5 \\
Combined Uncertainty &1.3\\
 & \\
 Difference (Glass-Open) & -0.5 \\
 Total Uncertainty & 2.5\\
\bottomrule
\end{tabular}
\end{center}
\end{table}

\eject
\section{Experimental Results II: New Software}

The data collected by dropping both solid and open retroreflectors, described in Sect. 5, was also processed using software based on the relativistic treatment using Eq. (\ref{finalmodel}), above.  Frequency modulation of the interference signal is represented by including the constants $A$ and $B$, as derived in Sect. 3;  this modulation theory includes both non-relativistic and relativistic effects arising from the last factor in Eq. (\ref{finalmodel}).  Frequency modulation coefficients suggested in \cite{niebauer95} would be equivalent to the replacement
\be\label{modreplace} 
\big(A \cos(\omega t)+B \sin(\omega t)\big)\bigg(F_{nr}+\frac{1}{c}F_{rel}\bigg) \rightarrow A \cos(\omega t)+B \sin(\omega t)\,.
\ee
RMS residuals using the present model are approximately 0.0024 fringes and are from 2\% to 4\% smaller than the replacement modulation scheme suggested in Eq. (\ref{modreplace}) . 

The data for each of the solid/open retroreflectors was processed in two ways.  One way consisted of combining both non-relativistic and relativistic effects, as expressed in Eq. (\ref{finalmodel}).  With this model the non-relativistic effects can be separated out by taking $c \rightarrow \infty$; fitting with this completely non-relativistic model, the relativistic corrections can be calculated using Eq. (\ref{relativisticcorrections}).  The differences between the fully relativistic model and the non-relativistic model for $g_t$ were found to be equal to the relativistic corrections to within $.02\ \mu$Gal.  In all cases the relativistic corrections were $-13.9\ \mu$Gal so a separate column for these corrections is not given in Table 3.  The results for the complete model including relativistic effects are given in Table 4.
 
\begin{table}
\caption{Experimental Results using software based on Eq. (\ref{finalmodel}).  Height is the distance from the bench mark to the optical center of the retroreflector at the top of the drop.   The gravity values are differenced from 979 622 000 $\mu$Gal.  The vertical gravity gradient is used to transfer the raw values to a common height of 139.6 cm above the bench mark. Chamber \# 1 labels the solid retroreflector.}
\vskip 8 pt
\begin{center}
\begin{tabular}{c c c c c c}
\toprule
Date & Chamber & Number of Drops & Height& Gravity & Uncertainty \\
& & Processed & (cm)&  ($\mu$Gal) & \\
\midrule
8.6.2019 & \# 1 & 1190 & 139.8 & 710.3 & 1.8\\
8.8.2019 & \# 1 & 983 & 139.8 & 709.6 & 1.8\\
8.8.2019 & \# 2 & 485 & 139.45 & 711.8 & 1.8\\
8.9.2019 & \# 1 & 496 & 139.75 & 710.7 &1.8\\
8.9.2019 & \# 2 & 489 & 139.5  & 710.4 & 1.8\\
8.9.2019 & \# 1 & 496 & 139.8  & 709.7 & 1.8\\
8.9.2019 & \# 2 & 493 & 139.45  & 711.9 & 1.8\\
\bottomrule
\end{tabular}
\end{center}
\end{table}

\begin{table}
\caption{Experimental results using software based on Eq. (\ref{finalmodel}). The mean gravity values are differenced from 979\ 622\ 000 $\mu$Gal.  All units are $\mu$Gal.}
\vskip 8 pt
\begin{center}
\begin{tabular}{c c}
\toprule
Mean Gravity Chamber \# 1 (Glass) & 710.1\\
\midrule
Error of the Mean &1.1\\
Systematic Uncertainty &0.5 \\
Combined Uncertainty & 1.2\\

\midrule
Mean Gravity Chamber \# 2 (Open) & 711.4 \\
\midrule
Error of the Mean &1.2\\
Systematic Uncertainty &0.5 \\
Combined Uncertainty & 1.3\\
 & \\
 Difference (Glass-Open) & -1.3 \\
 Total Uncertainty & 2.5\\
\bottomrule
\end{tabular}
\end{center}
\end{table}

\section{Conclusions}

Comparison of the values of gravity determined by dropping open and solid glass retroreflectors shows good agreement, to within the estimated uncertainties.  Statistical and systematic uncertainties are dependent on the gravimeter used for the measurements and are the same for retroreflectors of both types.  Although no valid error has been found in the expression in \cite{ashby18} for the phase difference between interference and test beams when combined at the beamsplitter, literal application of that difference to experimental data leads\cite{ashby18} to a prediction of several $\mu$Gal difference in $g$ between open and solid retroreflectors; this difference disappears when the ambiguity in the argument of a trigonometric function such as a cosine is accounted for.  The intensity of the interference fringe pattern contains constant terms: 
\be 
\vert\psi \vert^2 \approx \cos\bigg(\frac{2}{\lambda}{-(Z_0+D n-d)+....}\bigg)\,.
\ee
Whether open or solid, such terms cannot be measured because a constant $2 \pi M$, where $M$ is an arbitrary integer, can be added to the argument of the cosine reducing the argument to a polynomial in the time with negligible constant terms. 

An improved treatment of the effect of frequency modulation, with modulation index $\beta>>1$ has been given, implemented in software, and applied to data for both open and solid retroreflectors.  Typically for a given drop with a time series of more than 2400 times, each corresponding to a fringe spacing of 400 fringes, the rms residual between the specified number of fringes and the theoretical number, given by Eq. (\ref{finalmodel}), is 0.02 fringes.  Two methods for computing relativistic corrections have been implemented: (1) the difference between gravity determined by the fully relativistic model, and gravity determined by a completely non-relativistic model, consistently gives -13.88 $\mu$Gal in the present case.  This difference is accurately given by Eq. (\ref{relativisticcorrections}).  

\appendix
\section*{Appendix: Distance of fall vs. wavelength }

The phase difference between reference and test beams if recombined at the beamsplitter is given in Eq. (\ref{phasebs}).  The corresponding number of fringes is obtained by replacing $\Omega$ by $2\pi c/\lambda $ and dividing by $2 \pi$. The number of fringes is then
\be
N(t)=-\frac{2}{\lambda}(Z_{cm}(t)+D n-d)\bigg(1-\frac{V_{cm}(t)}{c}\bigg)+{\rm constant}\,.
\ee  

Suppose that during some small time interval the CM position and velocity change by $\Delta Z_{cm}$ and $\Delta V_{cm}$, respectively.  The number of fringes observed will be  
\be
\Delta N=-\frac{2}{\lambda}(\Delta Z_{cm})(\bigg(1-\frac{V_{cm}(t)}{c}\bigg)+\frac{2}{\lambda}(Z_{cm}(t)+D n-d)\bigg(\frac{\Delta V_{cm}(t)}{c}\bigg)\,.
\ee
If the retroreflector falls by a half-wavelength, $\Delta Z_{cm} = -\lambda/2$, then the number of fringes observed will be
\be
\Delta N=1-\frac{V_{cm}(t)}{c}+\frac{2}{c\lambda}(Z_{cm}(t)+D n-d)\Delta V_{cm}(t)\,.
\ee
and the number is unity, with relativistic corrections.

\end{document}